\documentclass{article}
\usepackage{spconf,amsmath,graphicx}
\usepackage{amssymb}
\usepackage{makecell}
\usepackage{pgfplots}
\usepackage{bm}
\usepackage{hyphenat}

\usepackage[absolute]{textpos}
\newcount\colveccount
\newcommand*\colvec[1]{
    \global\colveccount#1
    \begin{pmatrix}
        \colvecnext
    }
    \def\colvecnext#1{
        #1
        \global\advance\colveccount-1
        \ifnum\colveccount>0
        \\
        \expandafter\colvecnext
        \else
    \end{pmatrix}
    \fi
}

\DeclareMathOperator*{\grad}{\bm \nabla\!}
\newcommand{\norm}[1]{\left|#1\right|}

\newcommand{\shrinkplot}[4]{
\begin{tikzpicture}
\def\la{#1};
\def\lb{#2};

\begin{axis}[
    width=\axisdefaultwidth,
    height=0.75*\axisdefaultwidth,
    samples=100,
    domain=0:#3, 
    axis lines=left,
    ymin = 0,
    ymax = #4,
    xtick = {0, #3},
    ytick = {0, #4},
    xlabel = {$x$},
    ylabel = {$S(x)$},
    xlabel near ticks,
    ylabel near ticks
  ]
      
\addplot[very thick, blue] plot (\x,
                            { \x * (1 -(2 * exp(max(-25,-\x*\x / (\la * \la)))  
                                          - exp(max(-25,-\x*\x / (\lb * \lb)))))
                            });
                        
\addplot[thin, dashed, black] plot (\x,\x);
\end{axis}

\end{tikzpicture}
}

\title{Learning a Generic Adaptive Wavelet Shrinkage Function for Denoising}
\name{Tobias Alt and Joachim Weickert
      \thanks{This work has received funding from the European Research Council 
              (ERC) under the European Union's Horizon 2020 research and 
              innovation programme (grant agreement no. 741215, ERC Advanced 
              Grant INCOVID).}
     }

\address{Mathematical Image Analysis Group,
         Faculty of Mathematics and Computer Science,\\
         Campus E1.7, Saarland University, 66041 Saarb\"ucken, Germany.\\
         \{alt,weickert\}@mia.uni-saarland.de}

\begin{document}
%\ninept
%
\maketitle
\begin{abstract}
The rise of machine learning in image processing has created a gap 
between trainable data-driven and classical model-driven approaches: While 
learning-based models often show superior performance, classical ones are often 
more transparent. To reduce this gap, we introduce a generic wavelet shrinkage 
function for denoising which is adaptive to both the wavelet scales as well as 
the noise standard deviation. It is inferred from trained results of a tightly 
parametrised function which is inherited from nonlinear diffusion. Our proposed
shrinkage function is smooth and compact while only using two 
parameters. In contrast to many existing shrinkage functions, it is able to 
enhance image structures by amplifying wavelet coefficients. Experiments show 
that it outperforms classical shrinkage functions by a significant margin.
\end{abstract}

\section{Introduction}\label{sec_intro}
With the advent of machine learning, most of the state of the art solutions in 
image processing tasks have become data-driven: Many degrees of freedom allow a 
trainable model to adapt well to the input data. While yielding highly 
tailored solutions, results are often too complex for a rigorous analysis. In 
contrast to this, classical model-driven approaches provide a sound theoretical 
foundation but often cannot compete with their learning-based counterparts.

%\textbf{Our Contribution.}
The goal of this work is to combine the advantages of model- and data-driven 
approaches on the basis of wavelet shrinkage for denoising: Given a 
signal that has been corrupted by noise, the goal is to reconstruct 
the original image as well as possible by modifying parts of the signal in the 
wavelet domain. We exploit the flexibility of a learning-based approach to 
train a smooth shrinkage function that adapts both to the wavelet scales 
and to the noise level. The proposed shrinkage function can even amplify 
wavelet coefficients and thus enhance important image structures, a property 
that most established shrinkage functions do not share. A low number of 
trainable parameters allows us to manually inspect the learned results and 
infer smooth connections between them. From these, we design a generic compact 
shrinkage function that incorporates the learned adaptivity, while only using 
two parameters. Experiments show that our shrinkage function outperforms 
classical ones by a significant margin. To the best of our knowledge, this is 
the first work to present a learning-based shrinkage function with 
this level of compactness.

%\textbf{Related Work.}
Wavelet shrinkage has first been proposed by Donoho and Johnstone 
\cite{DJ94}. Therein, a noisy signal is transformed to a wavelet basis, 
resulting coefficients are modified by a shrinkage function, and then 
transformed back to obtain a denoised result. Classical 
shrinkage functions use a single threshold parameter for all scales of the 
wavelet transformation. However, individual scales might require distinct 
thresholds as they are differently affected by noise. To this end, adaptive 
shrinkage methods have been proposed. Early works of Zhang et al. \cite{ZD98, 
Zh01} already study a smooth adaptive shrinkage function with trainable 
thresholds. Other authors train arbitrarily shaped shrinkage functions 
\cite{HS08}, some also train the wavelets \cite{GP18} or even general adaptive 
filters for shrinkage operations \cite{SS14}. Non-trainable adaptive 
statistical models include \cite{CYM02, PSWS03}. Most learning-based 
methods produce large amounts of trained parameters while relations between 
them are rarely investigated. In contrast to this, we directly employ a 
tightly parametrised model from which it is easy to infer smooth underlying 
parameter relations. 

An important connection between 2D wavelet shrinkage and nonlinear diffusion 
filtering has been established by Mr\'azek and Weickert \cite{MW07}. It
allows us to directly translate a  
diffusivity into a trainable shrinkage function. We use a so-called 
\textit{Forward-and-Backward} (FAB) diffusivity, resulting in a shrinkage 
function that can amplify coefficients. Only few works employ this 
property directly \cite{SMCD03}, but also results for learned shrinkage 
functions suggest its usefulness \cite{HS08}. The corresponding concept 
of backward diffusion has also shown to be successful 
\cite{GSZ02a, CP16, WWG18}.

%\textbf{Organisation of the Paper.}
The remainder of this paper is structured as follows: In Section 
\ref{sec_classical} we review classical wavelet shrinkage. We propose our 
model in Section \ref{sec_ours}, which is experimentally evaluated in 
Section \ref{sec_experiments}. Finally, Section \ref{sec_conclusion} summarises 
our conclusions.

\section{Classical Wavelet Shrinkage}\label{sec_classical}
\subsection{Basic Concept}
Classical discrete wavelet shrinkage represents a noisy signal $\bm f = \bm v + 
\bm n$ in the wavelet basis, wherein the additive noise $\bm n$ is better 
separated from the true signal $\bm v$. This is achieved by the following 
three-step framework:
\begin{enumerate} 
    \item \textit{Analysis:} The input data $\bm f$ is transformed to 
    wavelet and scaling coefficients. In this representation, the noise affects 
    all wavelet coefficients while the signal is represented by only a few 
    significant ones \cite{Ma98}. 
    
    \item \textit{Shrinkage:} A shrinkage function $S_\theta$ with a 
    threshold parameter $\theta$ is applied individually to the wavelet 
    coefficients. The scaling coefficients remain unchanged.
    
    \item \textit{Synthesis:} The denoised version $\bm u$ of $\bm f$ is 
    obtained by back-transforming the modified wavelet coefficients.
\end{enumerate} 
Many shrinkage functions have been proposed. We will consider the most 
prominent ones of soft \cite{Do95}, hard \cite{Ma98}, and garrote \cite{Ga98} 
shrinkage for comparison. While these functions are easy to use in a practical 
setting, they suffer from applying the same threshold parameter to all 
scales and their binary decision structure. Finer scales might require a  
different threshold than coarser scales as they are more affected by 
noise. Furthermore, there is no clear separation between noise and signal 
coefficients such that eliminating too many coefficients always destroys signal 
details and eliminating too few leaves noise in the reconstruction. 

The classical wavelet transformation is not shift-invariant: Shifting the 
input signal $\bm f$ will produce a different set of wavelet coefficients. To 
overcome this problem, Coifman and Donoho suggested \textit{cycle spinning} 
\cite{CD95}: The input signal is shifted, wavelet shrinkage is applied, and the 
results are averaged for all possible shifts. This yields the shift-invariant 
but redundant \textit{non-decimating} wavelet transformation.

\subsection{Relation to Nonlinear Diffusion}
For the two-dimensional wavelet transform, wavelet coefficient channels $w_x,
w_y$, and $w_{xy}$ for $x$-, $y$-, and diagonal direction are obtained. To 
design rotationally invariant shrinkage rules, special care is required.
Mr\'azek and Weickert \cite{MW07} achieve this by a channel coupling that
is inspired by nonlinear diffusion filtering. Their Haar wavelet shrinkage 
rule is given by
\begin{equation}\label{eq_coupledshrinkage}
S_\theta\left(\!\!\colvec{3}{w_x}{w_y}{w_{xy}}\!\!\right)
= \Big(1 - g\left(w_x^2 + w_y^2 + 2 \, w_{xy}^2\right)\!\!\Big)
\colvec{3}{w_x}{w_y}{w_{xy}}\!.
\end{equation}
The argument $\,w_x^2 + w_y^2 + 2 \, w_{xy}^2\,$ is a consistent approximation
to the rotationally invariant expression $\norm{\grad u}^2$, where 
$\norm{\,.\,}$ is the $L^2$ norm and $\grad = \left(\partial_x, 
\partial_y\right)^\top$ denotes the spatial gradient operator.
The function $g$ is a \textit{diffusivity} function from a nonlinear diffusion 
filter \cite{PM90}. In nonlinear diffusion, filtered versions $u(\bm{x},t)$ of 
an image $f(\bm{x})$ are obtained by solving the partial differential equation
\begin{equation}
  \partial_t u = \text{div}(g(\norm{\grad u}^2) \grad u)
\end{equation}
with $u(\bm{x},0)=f(\bm{x})$ as initial condition and diffusion time $t$. 
The diffusivity $g$ steers the activity of the process. It is 
scalar-valued and becomes small at edges where $\norm{\grad u}^2$ is large. This
results in rotationally invariant edge-preserving denoising.

Mr\'azek and Weickert \cite{MW07} have shown that one explicit time step 
of nonlinear diffusion with diffusivity $g$ is equivalent to coupled Haar
wavelet shrinkage (\ref{eq_coupledshrinkage}). Thus, we can directly 
translate existing diffusivities into shrinkage functions. 

\section{Trainable Adaptive Wavelet Shrinkage}\label{sec_ours}
In our model, we equip the two-dimensional coupled Haar wavelet 
shrinkage approach from \cite{MW07} with a trainable adaptive shrinkage 
function. We sample a noisy image $f$ on a discrete grid with $2^L \times 2^L$ 
sampling positions and grid distance $h_x = h_y =: h$. The discrete image is 
represented by a vector $\bm f$. It is transformed with the non-decimating 
two-dimensional Haar wavelet transformation $\bm W$ into lowpass scaling 
coefficients at scale $L$ denoted by $\bm w^L$, and directional wavelet 
coefficients $\left(\bm w_x^\ell, \bm w_y^\ell,\bm 
w_{xy}^\ell\right)_{\ell=1}^L$ for different scales. To ensure that  
coefficients have the same range on all scales, we rescale the basis 
functions accordingly. A shrinkage function $S_{\bm\theta}$ with parameters 
$\bm \theta = \left(\bm \theta^1, \dots, \bm \theta^L\right)$ for each scale is 
applied to the wavelet coefficients. Coefficients on scale $\ell$ are modified 
component-wise by $S_{\bm\theta^\ell}$ according to the coupled shrinkage rule 
(\ref{eq_coupledshrinkage}). We obtain the reconstruction $\bm u$ by 
applying the backward transformation $\bm {\tilde W}$ to the modified set of 
wavelet coefficients and the unaltered scaling coefficients:
\begin{equation}\label{eq_our_shrinkage_model}
\bm u = \bm {\tilde W} S_{\bm\theta}\left(\bm W \bm f\right).
\end{equation}

\subsection{Choice of Shrinkage Function}
We found the Forward-and-Backward (FAB) diffusivity of Smolka \cite{Sm02} to be 
a good candidate for modelling our shrinkage function. It uses two 
\textit{contrast parameters} $\lambda_1$ and $\lambda_2$ that control the 
amount of forward and backward diffusion. The diffusivity is given by
\begin{equation}\label{eq_diff_proposed}
    g(s^2) = 2 \, \exp\left(\frac{-s^2}{\lambda_1^2}\right)
    -     \exp\left(\frac{-s^2}{\lambda_2^2}\right),
    \quad \lambda_2 \geq \lambda_1.
\end{equation}
It is translated into the shrinkage function according to the coupled 
shrinkage rule (\ref{eq_coupledshrinkage}). For the extreme case of $\lambda_1 
= \lambda_2$, the diffusivity simplifies to the exponential Perona-Malik 
diffusivity \cite{PM90}, corresponding to pure shrinkage. For larger 
differences between $\lambda_2$ and $\lambda_1$, the backward diffusion becomes 
more pronounced and the shrinkage function damps small coefficients and 
amplifies larger ones.

\subsection{Learning Framework} 
To train the shrinkage functions, we add Gaussian noise to a database of ground 
truth images $\left(\bm v_k\right)_{k=1}^{K}$. This yields noisy 
images $\bm f_k$ from which we compute denoised results $\bm u_k$. The 
trainable parameters for the proposed shrinkage function on scale $\ell$ are 
given by $\bm\theta^\ell = \left(\lambda_1^\ell, \lambda_2^\ell\right)$. As an 
objective function, we choose the mean square error between the reconstruction 
$\bm u_k$ and the corresponding ground truth image $\bm v_k$, averaged over all 
image pairs and normalised by the number of pixels:
\begin{equation}
    E\left(\bm\theta\right)
    = \frac{1}{K}\sum_{k=1}^{K}\frac{\lVert\bm u_k-\bm 
            v_k\rVert_2^2}{2^{2L}}.
\end{equation} 
The parameters are trained with the gradient-based L-BFGS algorithm 
\cite{LN89}. 
To that end, we compute the gradients of the objective function w.r.t.~all 
trainable parameters.

\section{Experiments}\label{sec_experiments}
In our experimental setup, we use $400$ images from the BSDS500 database 
\cite{AMFM11} as a training set and the $68$ images introduced in 
\cite{RB05} as a test set. Their grey values are in $\left[0,255\right]$ and 
the grid size is set to $h = 1$. From each image we select random regions of 
size $256 \times 256$, i.e. the number of scales is $L=8$. All images are 
corrupted by Gaussian noise of mean 0 and standard deviation $\sigma$. We do 
not clamp  the resulting pixel grey values to the original grey value range to 
preserve the Gaussian statistics of the noise. We have found the learned 
parameters to be robust w.r.t.~any reasonable initialisation, so no 
pretraining is performed.

\subsection{Evaluation of the Learned Shrinkage Functions}
\begin{figure}
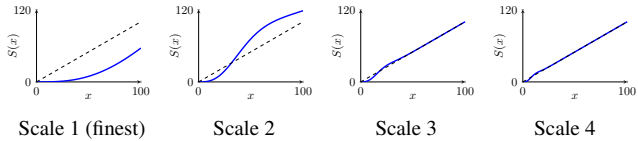

    \centering

\begingroup
\tikzset{every picture/.style={scale=0.45}}
\centering
\setlength{\tabcolsep}{1pt}
    \begin{tabular}{cccc}
        
          \shrinkplot{110.03639}{110.14656}{100}{120}
        & \shrinkplot{33.83090}{76.80491}{100}{120}
        & \shrinkplot{14.87020}{20.29808}{100}{120}
        & \shrinkplot{7.50205}{10.85179}{100}{120}\\
        
          \footnotesize Scale 1 (finest)
        & \footnotesize Scale 2 
        & \footnotesize Scale 3
        & \footnotesize Scale 4
        
    \end{tabular}

\endgroup

    \caption{Trained shrinkage functions for $\sigma = 25$. On 
             fine scales, both shrinkage and amplification are performed. The 
             coarser the scale, the less wavelet coefficients are modified.}
    
    \label{fig_function_inspection}
\end{figure}

In a first experiment, we train the adaptive shrinkage function for $\sigma = 
25$. The learned shrinkage functions for different scales are presented in 
Figure \ref{fig_function_inspection}. On the first and finest wavelet scale, 
all coefficients are shrunken. On the second scale, we observe coefficient 
amplification. We presume that this compensates the loss of image details 
caused by shrinking coefficients on the first scale. All further scales do not 
perform significant shrinkage as the learned function approaches the identity. 
Therefore, we do not display coarser scales $\ell > 4$.

\begin{figure}
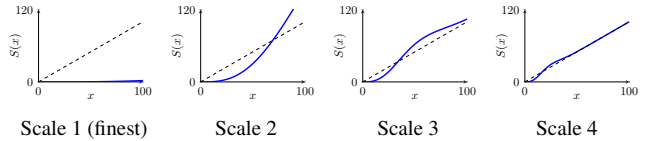

    \centering
    
\begingroup
\tikzset{every picture/.style={scale=0.45}}
\centering
\setlength{\tabcolsep}{1pt}
    \begin{tabular}{cccc}
        
          \shrinkplot{941.18963}{488520.93550}{100}{120}
        & \shrinkplot{82.99902}{2167061.40688}{100}{120}
        & \shrinkplot{33.01378}{56.59767}{100}{120}
        & \shrinkplot{14.92936}{22.32976}{100}{120}\\
        
          \footnotesize Scale 1 (finest)
        & \footnotesize Scale 2 
        & \footnotesize Scale 3
        & \footnotesize Scale 4
        
    \end{tabular}

\endgroup

    \caption{Trained shrinkage functions for $\sigma = 50$. With increasing 
             noise, shrinkage becomes more drastic and more scales are 
             involved.}
    
    \label{fig_function_inspection2}
\end{figure}

When we increase the noise level to $\sigma = 50$, we observe that more 
scales are involved in the shrinkage process. The trained shrinkage functions 
are displayed in Figure \ref{fig_function_inspection2}. Both configurations are 
in line with our conjecture that the diffusivity should change smoothly over 
the scales and the noise levels. Shrinkage and amplification decrease for 
coarser scales, i.e. $\lambda_1^\ell$ and $\lambda_2^\ell$ tend to $0$. With 
increasing noise, shrinkage and amplification become stronger and affect more 
scales. 

\subsection{Ablation Study}
To investigate the effectiveness of different aspects of our model, we 
perform the following ablation study. We start with classical hard wavelet 
shrinkage and equip it with the non-decimating wavelet transformation and the 
coupled shrinkage rule to enable a fair comparison. For $\sigma = 20$, we 
obtain an average PSNR on the test set of $27.88$dB. In a second step, we use 
the proposed shrinkage function restricted to $\lambda_2 = \lambda_1$, so no 
amplification can take place. Still, the shrinkage function does not adapt to 
the individual scales. This yields a comparable PSNR of $27.83$dB, showing that 
smoothness of the shrinkage function alone does not matter for reconstruction 
quality. When we remove the restriction on the shrinkage function, the PSNR 
increases to $28.06$dB which indicates that the amplification of wavelet 
coefficients is helpful for a good reconstruction. Finally, introducing 
adaptivity to the scales boosts the PSNR to $28.55$dB, demonstrating that the 
scale dynamic is the crucial ingredient for a good denoising result.

\subsection{Finding a Generalised Shrinkage Function}
\begin{figure}
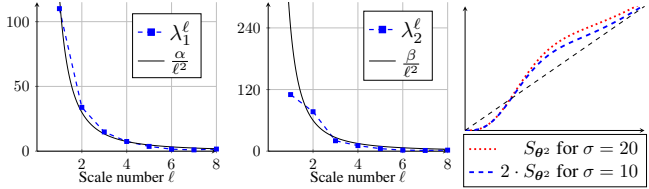

    \centering
    \begingroup
\tikzset{every picture/.style={scale=0.59}}
\centering
\setlength{\tabcolsep}{0pt}
    \begin{tabular}{ccc}
        
         \input{figures/fig_param_inspection/l1_over_levels_25.tex} 
        &\input{figures/fig_param_inspection/l2_over_levels_25.tex}
    
        &
        \begin{tikzpicture}
        
        \begin{axis}[
        width=\axisdefaultwidth,
        height=0.75*\axisdefaultwidth,
        samples=200,
        domain=0:100, 
        axis lines=left,
        ymin = 0,
        ymax = 100,
        ticks = none,
        legend entries={$S_{\bm\theta^2}$ for $\sigma = 20$,
                        $2 \cdot S_{\bm\theta^2}$ for $\sigma = 10$}, 
        legend cell align=right,
        legend style={at={(0.85,-0.05)},anchor=north, nodes={scale=1.2}}
        ]
        
        %second level of sigma_n = 20
        \def\la{26.32138}
        \def\lb{55.11412}
        
        % the max() formulation is a workaround for arithmetic overflow if la, 
        % lb are too small and x gets too large. It simply replaces something 
        % tending towards e^-infty by e^-25, which should be close enough. 
        \addplot[very thick, dotted, red] 
            plot (\x,{\x * (1 -(2 * exp(max(-25,-\x*\x / (\la * \la)))  
                                  - exp(max(-25,-\x*\x / (\lb * \lb)))))
                     });
        
        %second level of sigma_n = 10 scaled
        \def\la{2*12.94778}
        \def\lb{2*23.78630}
        \addplot[very thick, dashed, blue] 
            plot (\x,{\x * (1 -(2 * exp(max(-25,-\x*\x / (\la * \la)))  
                                  - exp(max(-25,-\x*\x / (\lb * \lb)))))
                     });
        
        \addplot[thin, dashed, black] plot (\x,\x);
        \end{axis}
        
        \end{tikzpicture}
    \end{tabular}

\endgroup
    
    \caption{Relations between trained parameters $\lambda_1^\ell, 			
             \lambda_2^\ell$ and scale $\ell$ (left, middle) and noise level 
             $\sigma$ (right).}
    
    \label{fig_param_inspection}
\end{figure}

So far, the contrast parameters are trained for each pair of scale $\ell$ and 
noise level $\sigma$ from which we will now infer a generic 
relation. Figure \ref{fig_param_inspection} shows the evolution of both 
contrast parameters over the scales and the noise levels.
A function of type $\frac{\alpha}{\ell^2}$ with an appropriate 
scalar $\alpha$ can provide a good description of the scale dependence
of $\lambda_1^\ell$. For $\lambda_2^\ell$, the values on fine scales 
do not follow this relation, but in these cases all relevant 
coefficients are already covered by shrinkage through a large $\lambda_1$, 
making the choice of $\lambda_2$ irrelevant.

Regarding the relationship between the shrinkage functions and the noise 
standard deviation it was already noted in \cite{HS08} that a simple rescaling 
of shrinkage functions is sufficient for adapting to a new noise level. For our 
parametrisation, rescaling the complete shrinkage function is equivalent to 
rescaling both $\lambda_1$ and $\lambda_2$. In Figure 
\ref{fig_param_inspection} we can see that indeed such a rescaling is learned.

These two insights suggest that a suitable generalisation
of the shrinkage function parameters which is smooth over the 
scales $\ell$ and the noise standard deviation $\sigma$ is given by  
$\lambda_1(\ell,\sigma) = \frac{\alpha\sigma}{\ell^2}$ and 
$\lambda_2(\ell,\sigma) = \frac{\beta\sigma}{\ell^2}$ 
where $\alpha$ and $\beta$ are scalars that have yet to be determined. To 
empirically show that this parametrisation indeed captures the 
underlying relations in a reasonable way, we compare two models: One model 
trains the proposed shrinkage function for each pair $\left(\ell, 
\sigma\right)$ individually, while the other one only optimises the factors 
$\alpha$ and $\beta$ of the generalised parameters. To ensure a fair 
comparison, both models are trained on a new training and test set combining 
images with noise levels between $\sigma = 10$ and $\sigma = 60$ in steps of 
$2.5$. Indeed, the generic model performs only $0.3$dB worse than the model 
with individual parameters in terms of PSNR, while training only $2$ instead of 
$336$ parameters. With this result we conclude that the generic shrinkage 
function captures the adaptivity to scales and noise levels well. The 
scalars are learned as $\alpha = 5.4$ and $\beta = 8.9$, yielding a combined 
generic coupled shrinkage function (\ref{eq_coupledshrinkage}) with diffusivity
\begin{equation}\label{eq_shrinkfunction_final}
    g(s^2,\ell,\sigma) =  2 \exp\left(\frac{-s^2}{\left(
                                      \frac{5.4\, \sigma}{\ell^2}
                                      \right)^2}\right)
                         -  \exp\left(\frac{-s^2}{\left(
                                      \frac{8.9\, \sigma}{\ell^2}
                                      \right)^2}\right).
\end{equation}

\subsection{Comparison to Classical Shrinkage}
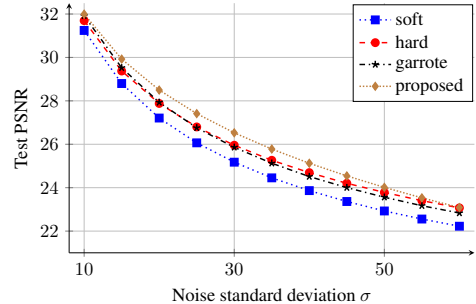
\begin{figure}
    \centering
    \begin{tikzpicture}[scale = 0.7]

\begin{axis}
[
width=1.1*\axisdefaultwidth,
height=0.75*\axisdefaultwidth,
samples=500,
axis lines=left,
ymin = 21, ymax = 32.5,
xmin = 8, xmax = 62,
xtick = {10, 30, 50},
ytick = {22, 24, ...,32},
ylabel={Test PSNR},
xlabel={Noise standard deviation $\sigma$},
xlabel near ticks,
ylabel near ticks,
grid = major,
legend entries={soft,
                hard,
                garrote,
                proposed},
legend cell align=left,
legend style={at={(0.7, 0.8)}, anchor = west}
]

\addplot+[thick, dotted, mark options={solid}, blue, mark = square*]
table[x = Noise, y = PSNR] 
{figures/fig_quality_comparison/soft.dat};

\addplot+[thick, dashed, mark options={solid}, red, mark = *]
table[x = Noise, y = PSNR] 
{figures/fig_quality_comparison/hard.dat};

\addplot+[thick, dashdotted, mark options={solid}, black, mark = star]
table[x = Noise, y = PSNR] 
{figures/fig_quality_comparison/garrote.dat};

\addplot+[thick, densely dotted, mark options={solid}, brown, mark = diamond*]
table[x = Noise, y = PSNR] 
{figures/fig_quality_comparison/ours.dat};

\end{axis}

\end{tikzpicture}
    
    \caption{PSNR values on the test set (higher is better) for individually 
             optimised classical approaches and our generic shrinkage function. 
             The proposed function outperforms all classical shrinkage 
             functions.}
    
    \label{fig_quality_comparison}
\end{figure}

Lastly, we compare our generic shrinkage function to soft, hard, and garrote 
shrinkage over a range of noise levels. We optimise the threshold parameter of 
the classical functions individually for each noise level, while the generic 
function is used as is from (\ref{eq_shrinkfunction_final}). The results are 
displayed in Figure \ref{fig_quality_comparison}. Although the classical 
approaches are optimised for each noise level, they are inferior to the generic 
shrinkage function. Over the range of noise levels used for training, 
improvements of up to $0.65$dB with an average of $0.34$dB are obtained 
compared to the best classical approaches.

\begin{figure}
    \centering
    
\begingroup
\centering
\setlength{\tabcolsep}{1pt}
\begin{tabular}{ccc}    
     \footnotesize\makecell{noisy image, \\ PSNR $14.74$dB}
    &\footnotesize\makecell{soft, \\ PSNR $21.62$dB}
    &\footnotesize\makecell{hard, \\ PSNR $22.19$dB} 
    
    \\
    
    \includegraphics[width =  0.31\linewidth] 
    {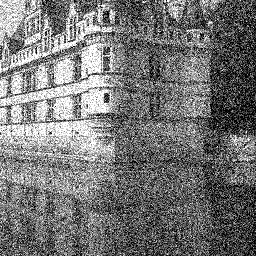}
    &\includegraphics[width = 0.31\linewidth]
    {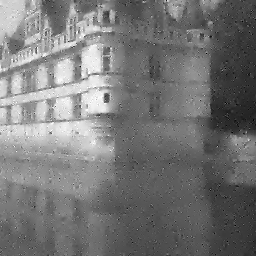}
    &\includegraphics[width =  0.31\linewidth] 
    {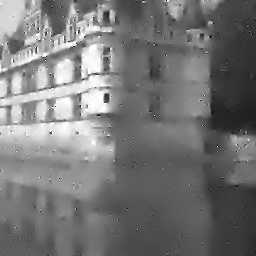}
    
    \\
    
    \includegraphics[width = 0.31\linewidth]
    {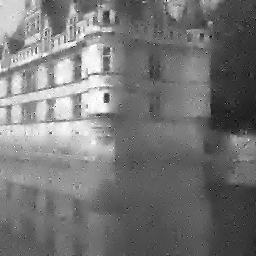}
    &\includegraphics[width = 0.31\linewidth]
    {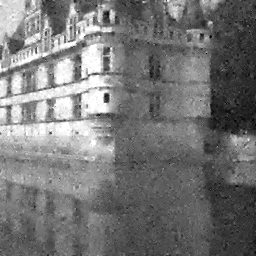}
    &\includegraphics[width = 0.31\linewidth]
    {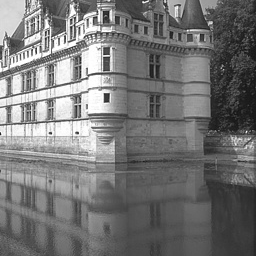}
    
    \\
    
      \footnotesize\makecell{garrote, \\ PSNR $22.10$dB}
    & \footnotesize\makecell{proposed, \\ PSNR $22.88$dB}
    & \footnotesize ground truth

\end{tabular}

\endgroup

    \caption{Comparison the results of classical shrinkage 
        functions and the generic shrinkage function for $\sigma = 50$. For the 
        generic function, the image is significantly less blurred.}
    
    \label{fig_image_comparison}
\end{figure}

For $\sigma = 50$, Figure \ref{fig_image_comparison} shows reconstructions 
along with the noisy input and the ground truth image. Soft shrinkage 
blurs the image too strongly since all wavelet coefficients are decreased by 
the same margin. Hard shrinkage suffers from remaining noise as it does not 
shrink large noisy wavelet coefficients. While less pronounced, this is 
also the case for garrote shrinkage. Both garrote and hard shrinkage also blur 
important image structures. Our generic shrinkage function outperforms all 
classical approaches. By strongly shrinking coefficients on 
fine scales, noise is efficiently removed. To compensate for lost image 
details, amplification of wavelet coefficients on coarser scales 
enhances important structures.

\section{Conclusions}\label{sec_conclusion}
Our approach of learning a compact shrinkage function for wavelet 
denoising combines the advantages of model-driven and data-driven 
approaches: In contrast to other parameter learning strategies, we
can cope with as little as two parameters without substantially
sacrificing performance. This results in an interpretable shrinkage 
function and a transparent, but adaptive model.

In our ongoing work we extend these findings to other adaptive 
nonlinear approaches such as diffusion evolutions.  


\bibliographystyle{IEEEbib}
\bibliography{myrefs}

\end{document}